\documentclass[fleqn,12pt,twoside]{article}
\usepackage{espcrc1}

\usepackage{graphicx}

\title{Negative Elliptic Flow from Anomaly Induced DCC Formation}

\author{Masayuki Asakawa\address{Department of Physics, Kyoto University,
Kyoto, 606-8502, Japan}, 
Hisakazu Minakata\address{Department of Physics, 
Tokyo Metropolitan University, 
Tokyo 192-0397, Japan}\thanks{Based on talk presented at 
XVI International Conference on Particles and Nuclei (PaNic02), Osaka, 
Japan, September 30-October 4, 2002.},
Berndt M\"uller\address{Department of Physics, Duke University,
Durham, NC 27708-0305, U.S.A.}}

\begin{document}

\maketitle

\begin{abstract}
We discuss characteristic experimental signatures related to 
the mechanism of DCC formation triggered by the chiral U(1) anomaly 
in relativistic heavy ion collisions. 
We predict an enhancement of the fraction of neutral pions compared 
with all pions in the direction perpendicular to the scattering 
plane. To quantify the effect on the angular distribution of neutral
pions, we compute the elliptic flow parameter $v_2$ as a function of 
the transverse momentum. We find values of order $-0.05$ at small 
momenta for neutral pions.
We also compute the $v_2$ parameter for inclusive photons, which is 
easier to measure, and confirmed that the negative a few percent 
effect prevails in this observable.

\end{abstract}

\section{Introduction}

Some time ago we predicted that the chiral U(1) anomaly may 
provide an efficient trigger mechanism for the formation of 
coherent domains of pion fields, the disoriented chiral 
condensate (DCC), in relativistic heavy ion collisions \cite{MM96}.
By implementing the anomaly effect into the linear sigma model 
simulation code developed by Asakawa et al. \cite{AHW95}, 
we demonstrated that this mechanism indeed enhances the DCC 
formation \cite{AMM98}.

In the following section, we present a compact summary of this idea. 
(Those who are familiar with our idea can skip the rest of the section 
and can go directly to the section 2.)
It is natural to suspect a possible significance of the electromagnetic
interaction in DCC formation because it breaks isospin symmetry, and 
DCC is the phenomenon of formation of coherent domains with definite 
isospin. In relativistic heavy ion collisions there exist transient 
strong electromagnetic fields, which could affect the formation of 
chiral domains. This idea provided our starting point of the subsequent 
development.

We quickly recognized that a time-independent and spatially 
uniform electromagnetic fields do not affect the chiral orientation 
of the ground state of the linear sigma model at the one-loop level. 
Instead, we found that a coherent effect can arise from 
electromagnetism via the chiral anomaly \cite{MM96}.

We have formulated the anomaly effect as an initial ``kick'' to 
the $\pi^0$ fields, because the collision time scale is much 
shorter than the time required for DCC formation. 
In the framework of ref.~\cite{MM96}, however, it was difficult 
to assess how effective the ``kick'' is;
it is a small effect in magnitude, but it is a coherent effect 
which extends over the nuclear dimension. So we needed some 
powerful tool to analyze the intricate question of effectiveness
of the kick.
We found that the linear sigma model simulation is appropriate 
for this purpose. 
We have adopted the simulation code developed by Asakawa et al. 
\cite{AHW95}, rather than the original one used by Rajagopal and 
Wilczek \cite{RW93}, because we wanted to be as realistic as 
possible. For example, the newer code takes into account the 
boost invariant longitudinal expansion of the hadronic debris.

We found, to our surprise, that the anomaly kick is efficient 
in triggering a coherent motion of the chiral order parameter
and resulting in the formation of a domain of DCC \cite{AMM98}. 
The effect of the kick survives the non-linear evolution 
of the system and the coherent motion triggered by the kick 
continues until times of order $\sim$ 10 fm 
(see Figs.~10 and 11 of ref.~\cite{AMM98}). 
We need to emphasize that the persistence of the kick effect
may be, in part, due to the neglect of quantum dissipation 
in the classical simulation. We refer to ref.~\cite{AMM98} 
for further details.

To facilitate the discussion of experimental signatures of the 
anomaly mechanism in the next section we note here the crucial 
characteristic property of the anomaly kick. That is, 
the kick is proportional to the angular momentum of the two ions 
in collision and depends on the momentum direction perpendicular 
to the scattering plane. 
It points into the $I_3$ direction in the isospin space.
These properties derive from the vector structure of the anomaly
${\vec E}\!\cdot\!{\vec B}$ produced by two colliding ions 
\cite{AMM98}.

\section{Experimental signatures}

We now discuss possible characteristic experimental signatures 
of the anomaly-induced DCC formation in relativistic heavy ion 
collisions. We first summarize our work done previously, and 
then come to our new results.
We examine three types of observables, which we discuss one by one 
below. They are: 
(1) number ratio of $\pi^0$ to all $\pi$ \cite{AMM98}, 
(2) azimuthal angle distribution of the ratio of the number of 
$\pi^0$ to all pions and photons \cite{AMM02}, and 
(3) the elliptic flow parameters of  $\pi^0$ and photons.
We report (3) for the first time in this article. 
We have found that the latter two types of observables are 
particularly suitable for this purpose. 

In calculating experimental observables we use three values of 
the anomaly kick parameter $a_n = 0.05, 0.1,$ and $0.2$, 
and employ the coherent state approximation to extract particle 
number from field configurations of linear sigma model fields. 
Based on our previous estimate of the kick parameter at  
RHIC \cite{AMM98},
we focus on $a_n = 0.1$ to estimate the effect that would 
be seen in relativistic heavy ion collisions at RHIC.
See \cite{AMM98} for the definition of the anomaly kick parameter 
$a_n$ and more details of our calculational procedure.

\subsection{Number ratio of $\pi^0$ to all $\pi$}

Since the anomaly kick acts only on the $\pi^0$ field, it is natural 
to expect that neutral pions behave differently from charged pions 
in our mechanism. Therefore, we first computed the number ratio 
$n(\pi^0)/n(\mbox{all}~\pi$'$s)$ and found that the ratio exceeds 1/3 
(the value expected by isospin invariance) by about 13\% at low 
momenta, $|\vec{k}| \leq 250$ MeV \cite{AMM98}. 

\subsection{Azimuthal angle distributions of number ratio of 
$\pi^0$ to all $\pi$ and photons}

We then searched for other, clearer signatures, which would
unambiguously indicate the anomaly-induced DCC formation. 
Since the characteristic feature of the kick is that it has
a momentum dependence
perpendicular to the scattering plane, it is natural to study 
whether the number ratio of $\pi^0$ to all $\pi$ exhibits an
angular dependence correlated with the scattering plane. 
Thus, we examined the azimuthal angle distribution of number 
ratio of $\pi^0$ to all $\pi$. We found that the number ratio 
of $\pi^0$ to all $\pi$ shows a large variation with azimuthal 
angle, peaking in the direction perpendicular to the scattering 
plane. See Fig.~1. (Fig.3 of \cite{AMM02})

\begin{figure}[thb]
\begin{minipage}[t]{75mm}
\includegraphics[width=1.0 \linewidth]
{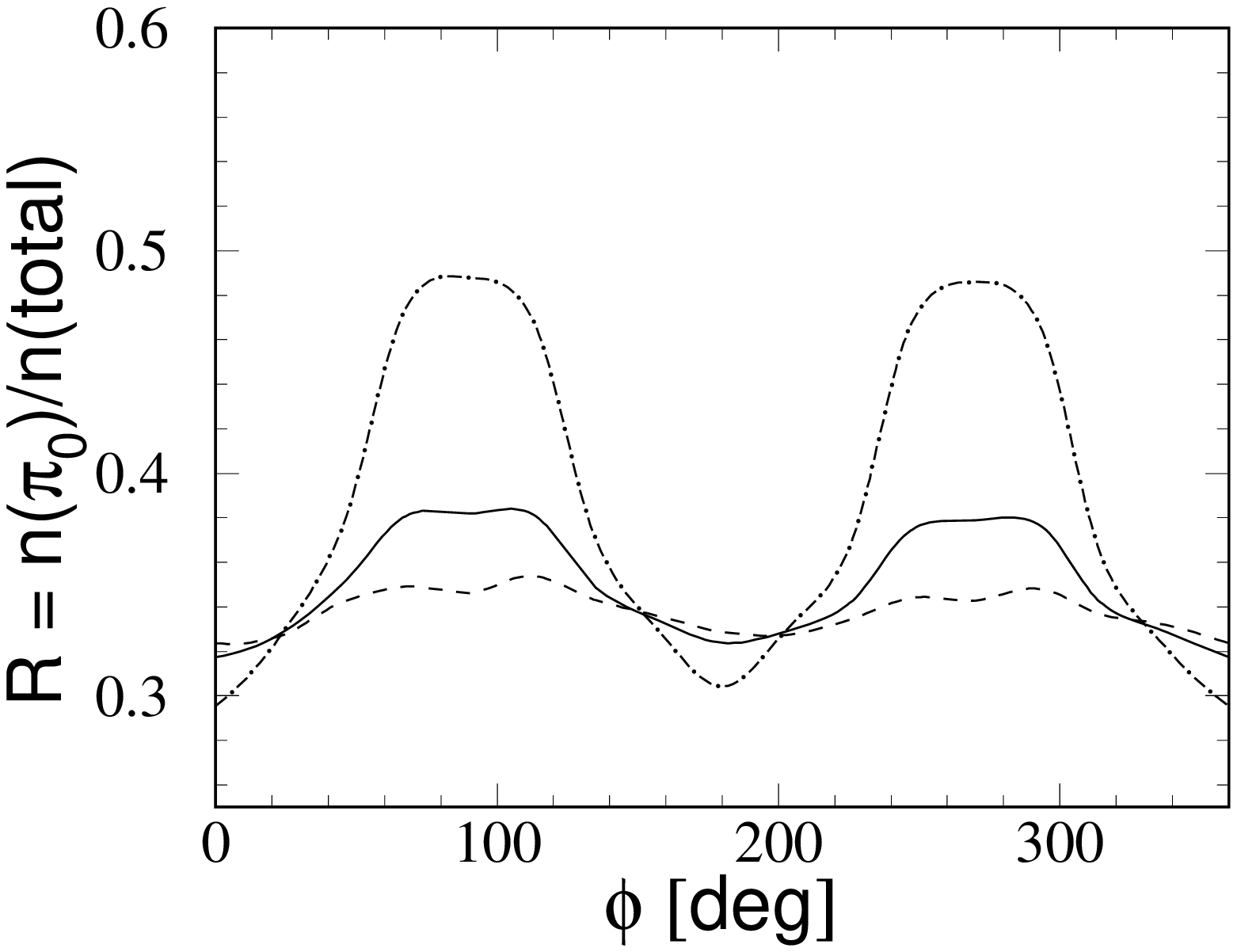}
\caption{
  Azimuthal angle $\phi$-dependence of the ratio
  $n_{\pi^0}/n_{\pi}$ for $R_0 =5$ fm at time $\tau = 11$ fm.
  Three lines indicate: 
  Dashed line: $a_n = 0.05$;
  solid line: $a_n = 0.1$;
  dash-dotted line: $a_n = 0.2$. 
  The two maxima lie in the direction normal to the scattering
  plane and are an expression of the spatial variation of the
  axial anomaly source term.}
\label{Zdep}
\end{minipage}
\hspace{\fill}
\begin{minipage}[t]{75mm}
\includegraphics[width=0.9 \linewidth]{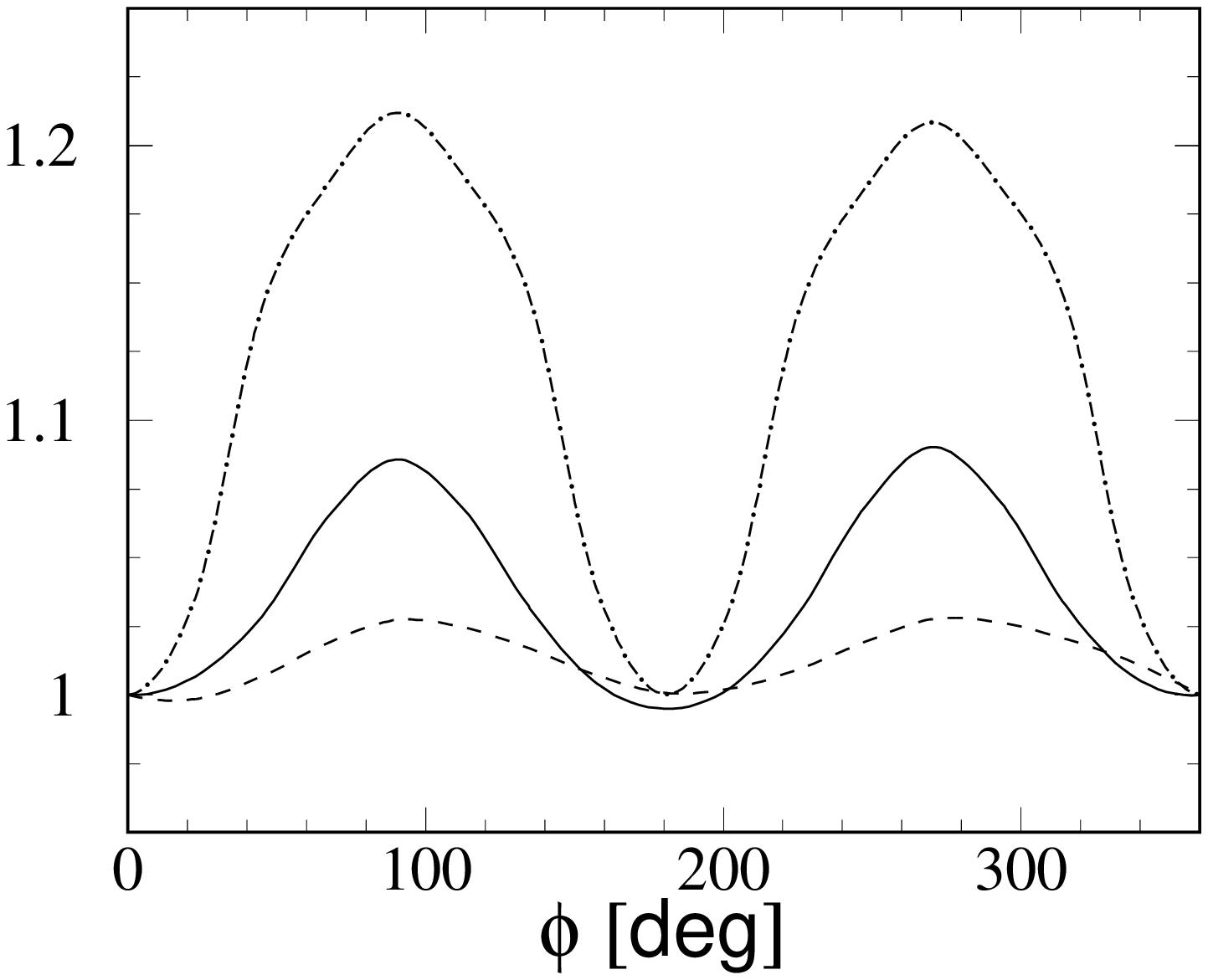}
\caption{
  Azimuthal angle $\phi$-dependence of the inclusive photon 
  distribution in the range $0 < k_T < 500$ MeV. 
  The magnitude of the anomaly kick $a_{n}$ is taken as in 
  Fig.~\protect{\ref{Zdep}} and displayed by the same symbols.
  The distribution is normalized to one at $\phi=0$.}
\label{gammas}
\end{minipage}
\end{figure}

An important question is whether this azimuthal asymmetry 
persists in the inclusive photon distribution, which is much 
easier to measure. We have examined this possibility and 
obtained a very encouraging result: photons exhibit an azimuthal 
angle variation of about 10\%, even if we include a rather wide 
range of photon transverse momenta $k_T \leq 500$ MeV. 
See Fig.~2 (Fig.4 of \cite{AMM02}). If it exists, the effect should 
be observable in the RHIC experiments.

\subsection{Elliptic flow parameter of $\pi^0$ and photons}

It has become an industry to measure the elliptic flow 
parameter $v_2$ \cite{flow,olli} in heavy ion collisions. 
Of course, the main motivation for the enthusiasm comes from
elucidating hydrodynamical behavior of the hadronic matter.
But since our anomaly-induced mechanism of DCC formation also 
predicts an anisotropic flow of $\pi^0$, it is quite natural to 
convert our prediction into the language of elliptic flow. 

We have run a Monte-Carlo code coupled with the Asakawa et al. 
sigma model simulation code to compute the elliptic flow parameter 
$v_2$ for neutral pions and inclusive photons from $\pi^0$ decay.
Since the computation of the photon distribution is rather 
time consuming, we define an approximate $v_2$ as follows:
\begin{equation}
v_2^{app}(k_{T}) = 
\frac{(n_{\parallel}-n_{\perp})}
     {\frac{2}{\pi}(n_{\parallel}+n_{\perp})},
\end{equation}
where, $n_{\parallel}$ denotes the number of $\pi^0$ or gammas 
emitted with azimuthal angle $\phi$ measured from the reaction 
plane in the range 
$-45 \deg \leq \phi <  45 \deg$, or 
$135 \deg \leq \phi < 225 \deg$, 
and
$n_{\perp}$ the number of $\pi^0$ or gammas 
in the range 
$45 \deg \leq \phi < 135 \deg$, or 
$225 \deg \leq \phi < 315 \deg$. 
This definition of $v_2$ is exact when the azimuthal asymmetry
has solely a quadrupole contribution.

We present in Fig.~3 the result of our computation of 
the approximate $v_2$ for $\pi^0$ and photons. 
The value of $v_2^{app}$ is $\sim 4-5$ \% for $\pi^0$ and 
$\sim$ 1 \% for photons. 
In the absence of other effects generating an elliptic flow, 
$v_2^{app}$ approaches unity at high momenta. 
The relativistic kinematics of $\pi^0 \rightarrow 2 \gamma$ decay 
is taken into account in the Monte-Carlo code and it is essential 
to produce this behavior.

\begin{figure}
\begin{minipage}[t]{75mm}
\centerline{\includegraphics[angle=-90,width=0.95\linewidth]{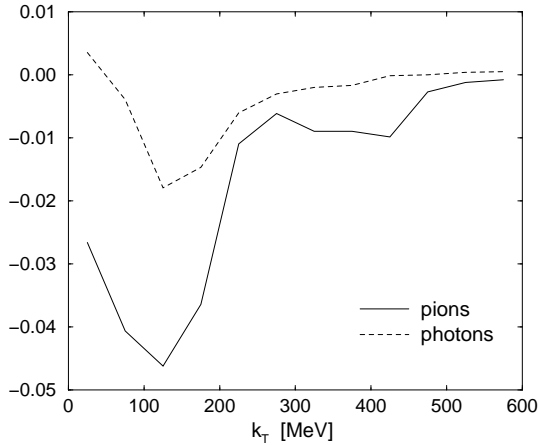}}
\end{minipage}
\hspace{5mm} 
\begin{minipage}[t]{75mm}
\vspace*{2.1cm}
\caption
{The approximate elliptic flow parameter $v_2^{app}$ 
defined in the text is presented for $\pi^0$ and $\gamma$. 
Both for $\pi^0$ and $\gamma$, $v_2^{app}$ is plotted at time 
$\tau = 11$ fm. The kick parameter is taken as $a_n = 0.1$.
}
\end{minipage}
\label{v_2}
\end{figure}

As we see in Fig.~3 $v_2^{app}$ for both $\pi^0$ and photons
are negative at low $k_T$, the behavior quite opposite to 
the one commonly observed. The latter behavior is usually 
attributed to the hydrodynamical expansion of hadronic debris. 
It would be interesting to see such a peculiar feature as 
negative $v_2$ in either $\pi^0$ or gamma distributions 
in heavy ion collisions. It may be the unique signature 
of DCC formation due to the anomaly mechanism.

Although they are small effects, 
we suspect that at least the photon $v_2$ is measurable. 
It may be difficult to measure $v_2$ for $\pi^0$, because 
one needs event-by-event reconstruction of $\pi^0$.
But for photons, measurement of either the inclusive single 
particle distribution with identification of the scattering plane, 
or the two particle azimuthal correlation would be enough.
Therefore, we urge RHIC experimentalists to try to measure 
the photon $v_2$ at low transverse momenta.


\end{document}